\documentstyle [twocolumn,epsf]{mn}
\newcommand{\mincir}{\raise
-3.truept\hbox{\rlap{\hbox{$\sim$}}\raise4.truept\hbox{$<$}\ }}
\newcommand{\magcir}{\raise
-3.truept\hbox{\rlap{\hbox{$\sim$}}\raise4.truept\hbox{$>$}\ }}
\newcommand{\minmag}{\raise
-3.truept\hbox{\rlap{\hbox{$<$}}\raise5.truept\hbox{$<$}\ }}
\newcommand{\be}{\begin{equation}}
\newcommand{\ee}{\end{equation}}
 \newcommand{\ba}{\begin{eqnarray}}
\newcommand{\ea}{\end{eqnarray}}
\newcommand{\brr}{\begin{array}}
\newcommand{\err}{\end{array}}
\newcommand{\bc}{\begin{center}}
\newcommand{\ec}{\end{center}}

\renewcommand{\baselinestretch}{1.0}

\title[Clustering and Bias of the Soft XMM point sources]
{The XMM-{\em Newton}/2dF Survey - VI: 
Clustering and Bias of the Soft X-ray Point Sources}
\author[Basilakos et al.]{S. Basilakos$^{1}$, M. Plionis$^{1,2}$, 
A. Georgakakis$^{1}$, \& I. Georgantopoulos$^{1}$. \\
\vspace{0.1cm}
$^1$ Institute of Astronomy \& Astrophysics, National Observatory of Athens, 
I. Metaxa \& V. Pavlou, Palaia Penteli, 15236 Athens, Greece \\
$^2$ Instituto Nacional de Astrofisica, Optica y Electronica (INAOE)
Apartado Postal 51 y 216, 72000, Puebla, Pue., Mexico
}

\begin{document}

\maketitle

\begin{abstract}
We study the clustering properties of X-ray sources detected in 
the wide area ($\sim 2 $deg$^{2}$) bright, contiguous 
XMM-{\em Newton}/2dF survey. We detect 432 objects  to a 
flux limit of $5\times 10^{-15}$ ergcm$^{-2}$s$^{-1}$
 in the soft 0.5-2 keV band. 
Performing the standard angular correlation function analysis, a
$\sim 3\sigma$ correlation signal between 0 and 150 arcsec is 
detected: $w(\theta<150^{''})\simeq 0.114\pm 0.037$. 
If the angular correlation function is modeled as a 
power law, $w(\theta)=(\theta_{\circ}/\theta)^{\gamma-1}$, 
then for its nominal slope of $\gamma=1.8$ we
estimate, after correcting for the integral constraint and the
amplification bias, that $\theta_{\circ}\simeq 10.4 \pm 1.9$ arcsec. 
Very similar results are obtained for the 462 sources 
detected in the total 0.5-8 keV 
band ($\theta_{\circ}\simeq 10.8 \pm 1.9$ arcsec).

%ig
Using a clustering evolution model which is constant in comoving
coordinates ($\epsilon=-1.2$), a luminosity dependent density evolution model for the
X-ray luminosity function and the concordance cosmological model ($\Omega_{\rm 
m}=1-\Omega_{\Lambda}=0.3$) we obtain, by inverting Limber's integral equation, 
a spatial correlation length of $r_{\circ}\sim 16$ $h^{-1}$
Mpc. This value is larger than that of previous {\sc ROSAT} surveys
as well as of the optical two-degree quasar redshift survey.
Only in models where the clustering remains constant in 
physical coordinates ($\epsilon=-3$), do we obtain an $r_{\circ}$
value ($\sim 7.5$ $h^{-1}$ Mpc) which is consistent with the above surveys.   
%ig
%which cover a similar redshift range
%($r_{\circ} \mincir 5\; h^{-1}$ Mpc; Carrera et al 1998).
%%%--- end Manolis

Finally, comparing the measured angular correlation function 
with the predictions of the concordance cosmological model, we find
for two different bias evolution models that 
the soft X-ray sources at the present time 
%%%--- start Manolis
should be biased with
respect to the underline matter fluctuation field with bias values
in the range (which depends on the biasing model used):
$1.9 \mincir b_{\circ} \mincir 2.7$ for $\epsilon=-1.2$ or
$1 \mincir b_{\circ} \mincir 1.6$ for $\epsilon=-3$.

\vspace{0.4cm}

\end{abstract}

\vspace{1.0cm}

\section{Introduction}
Active Galactic Nuclei (AGN) can be detected out to high redshifts and
therefore, study of their clustering properties can provide
information on both the large scale structure of the underlying matter
distribution and its evolution with redshift.  At optical wavelengths
the 2dF QSO redshift survey (2QZ; Croom et al. 2000) comprising over
25\,000 optically selected QSOs  in the range $z\approx0.3-3$ has
provided tight constraints on the spatial distribution of powerful
AGNs (Croom et al. 2001; Croom et al. 2002). A striking result from
this survey was that the  clustering properties of QSOs are comparable
to  those of local galaxies. Moreover, when studied as a function of
redshift the  clustering of these sources  was found to be constant
out to $z\approx3$.  

Optically selected AGN catalogues however, are believed to miss large 
numbers of dusty systems and therefore, provide a biased census of the
AGN phenomenon. X-ray surveys, are least affected by dust providing an
efficient tool for compiling uncensored AGN samples over a 
wide redshift range. From the
cosmological point of view an interesting question that remains to be
addressed is how the X-ray selected  AGNs trace the underlying mass
distribution and whether there are any differences with optically
selected samples. Despite the  importance of X-ray selected AGNs,
their clustering properties remain poorly constrained. Early studies
with the {\it Einstein} and the  {\it ROSAT} satellites have produced 
contradictory results. Boyle \& Mo (1993) used low redshift AGNs
detected in the Einstein Medium Sensitivity Survey (EMSS; reference)
and found only a marginally significant clustering signal at scales
$<10 \;h^{-1}$ Mpc. Vikhlinin \& Forman (1995) combined archival {\it
ROSAT} observations totaling $\rm 40\,deg^2$ and detected, for the
first time,  a statistically significant clustering signal using
angular correlation function  analysis. Their results suggest a
clustering length consistent with that of optically selected
QSOs. Akylas, Georgantopoulos \& Plionis (2000) used the {\em ROSAT}
All Sky Survey Bright Source Catalogue to explore the clustering of
nearby AGNs. They estimate $r_{\circ}=6.5\pm  1.0 \;h^{-1}$ Mpc, which
is also similar to nearby galaxies and the 2QZ survey
results. Contrary to the studies above that are based on an angular
correlation analysis, Carrera  et al. (1998) used redshift information
to measure the spatial correlation function of X-ray sources in the
{\it ROSAT} Deep  (Georgantopoulos et al. 1996) and RIXOS (Mason et
al. 2000) surveys. They detect only a marginally significant
clustering signal and argue that their results suggest that the X-ray
population is more weakly clustered than optically selected galaxies
or AGNs. Recently, Mullis et al. (2004) using the {\em ROSAT} 
North Ecliptic Pole (NEP) survey of relatively local 
X-ray selected AGNs, found a spatial 
correlation length of $r_{\circ} \simeq 7.4 \pm 1.8 \; h^{-1}$ Mpc
within the concordance cosmological model.

The new generation {\it Chandra} and XMM-{\it Newton}
telescopes have extended the studies above to the hard (2-8\,keV)
spectral band. Yang et al. (2003) used {\it Chandra} observations and
argued that hard (2-8\,keV) X-ray selected sources have large variance
(strong clustering) and are most likely associated with high density
regions. These authors also find that X-ray sources selected in the
soft (0.5-2\,keV) energy band are less clustered (about 1\,dex) than
hard ones. Recently, Basilakos et al. (2004) applied an angular 
correlation function
analysis to hard X-ray selected sources detected in the wide area,
shallow XMM-{\it Newton}/2dF survey. They find a strong signal
consistent with a spatial clustering length in the range  $r_{\circ}
\sim 10 -19 \; h^{-1}$\,Mpc (in the concordance cosmological model). This
also suggests that hard X-ray sources could trace the high density
peaks of the underlying mass distribution.  

In this paper we further explore the clustering properties of the
X-ray population exploiting the high sensitivity and  the large
field-of-view of the XMM-{\it Newton} observatory. In particular, we
extend the Basilakos et al. (2004) clustering study to sources
detected in the soft (0.5-2\,keV) and the total (0.5-8\,keV) spectral
bands of a wide area ($\rm \approx 2 \,deg^2$), contiguous XMM-{\it
Newton} survey (XMM-{\it Newton}/2dF survey). Our study provides the
first constraints on the clustering properties of the sources in the
above spectral bands using the XMM-{\it Newton}. Furthermore, we model
our X-ray source clustering and its evolution in an attempt to
derive their present time bias with respect to the underline mass
fluctuation field.

The structure of the paper is as follows. The X-ray sample is
presented in Section 2 and the angular correlation function analysis
is discussed in Section 3, while the spatial clustering predictions
are presented in section 4. Section 5 outlines the models used to 
interpret the angular correlation function results and the theoretical
interpretation of the X-ray source clustering.
Finally, we draw our conclusions in section 6. Hereafter and wherever
necessary we will assume the {\em concordance} cosmological model
(unless stated otherwise), ie.,
$\Omega_{\rm m}+\Omega_{\Lambda}=1$, $\Omega_{\rm m}=0.3$, 
$H_{\circ}=100 \;h $km s$^{-1}$ Mpc$^{-1}$ (Spergel et al. 2003; 
Tegmark et al. 2004)  with $h\simeq 0.7$ (Freedman et al. 2001; 
Peebles and Ratra 2002 and references therein) and baryonic density 
parameter $\Omega_{\rm b} h^2 \simeq 0.02$ (cf. 
Olive, Steigman \& Walker 2000; Kirkman et al 2003).

\section{The sample}
The X-ray data used in this study are from the XMM-{\it Newton}/2dF
survey. This is a shallow (2-10\,ksec per  pointing) survey carried
out by the XMM-{\it Newton} near the North Galactic Pole  [NGP;
RA(J2000)=$13^{\rm h}41^{\rm m}$;   Dec.(J2000)=$00\degr00\arcmin$]
and the South Galactic Pole [SGP; RA(J2000)=$00^{\rm h} 57^{\rm m}$,
Dec.(J2000)=$-28\degr 00\arcmin$] regions. A total of 18 XMM-{\it
Newton} pointings were observed equally split between the NGP and the
SGP areas. A number of pointings were discarded due to elevated
particle background at the time of  the observation  resulting in a
total of 13 usable XMM-{\it Newton} pointings. A full description of
the data reduction, source detection and flux estimation are presented
by Georgakakis et al. (2003, 2004). 
\begin{figure}
\mbox{\epsfxsize=8.4cm \epsffile{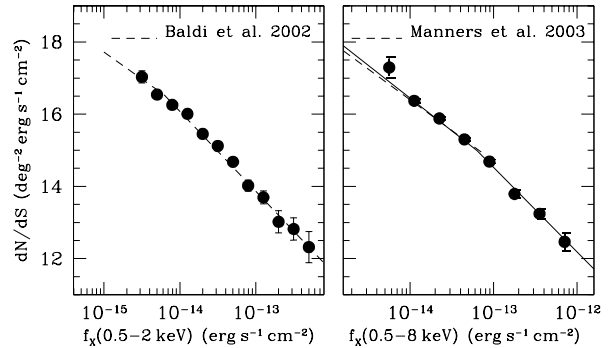}}
\caption{ {\sc Left panel}: The soft band (0.5-2\,keV) differential number
counts from the present survey in comparison with the best fit
double power-law to the counts from  Baldi et al. (2002; dashed
line). {\sc Right panel}: The total band (0.5-8\,keV) differential counts
from the present survey in comparison with the results from Manners et 
al. (2003; dashed line) spanning the flux range $\rm 10^{-15} - 8
\times 10^{-14} \, erg \, s^{-1} \, cm^{-2}$. At bright fluxes
extrapolation of the Manners et al. (2003) best fit relation
overestimates the observed X-ray source surface density. The
continuous line is our best fit relation to the 0.5-8\,keV counts from
the XMM-{\it Newton}/2dF survey using the double
power-law described in the text.}\label{fig_lognlogs}  
\end{figure}

Here we use the soft (0.5-2\,keV) and the total  (0.5-8\,keV) band
catalogues of the XMM-{\it Newton}/2dF survey. We only consider
sources at off-axis angles $<13.5$\,arcmin. The two  samples comprise
432 and 462 sources respectively above the 
$5\sigma$  detection threshold. The limiting fluxes are  $f_X(\rm
0.5-2)=2.7\times 10^{-15} \, erg \, \, s^{-1} cm^{-2}$ and $f_X(\rm
0.5-8)=6.0\times 10^{-15} \, erg \, \, s^{-1} cm^{-2}$. The
sensitivity of the XMM-{\it Newton} degrades from the center to 
the edge of the field of view (vignetting) and  therefore the limiting
flux varies across the surveyed area.  We account for this effect by
constructing sensitivity maps giving the area of the survey accessible
to point sources above a given flux limit. In the 0.5-8\,keV band
about 10 per cent  of the surveyed area is covered at the flux
$f_X(\rm 0.5 - 8 \,keV) =  10^{-14} \,erg \,s^{-1} \,cm^{-2}$. This
fraction increases to about 50 per cent  at $f_X(\rm 0.5 - 8 \,keV) =
2\times10^{-14} \,erg \,s^{-1} \,cm^{-2}$. In the soft band about 10
and 50 per cent of the total area is covered at the flux $f_X(\rm 0.5
- 2 \,keV) =  3.5 \times 10^{-15} \,erg \,s^{-1} \,cm^{-2}$ and $\rm
5\times10^{-15} \,erg \,s^{-1} \,cm^{-2}$ respectively. 

%ig
Unfortunately, the identification of our sources in the 0.5-2 keV
band remains largely unknown since optical spectroscopy
is not available for the large majority of them.
However, from other surveys in  the same band and 
of similar depth, we know that the vast majority 
of sources are associated with AGN. 
For example, among the 50 soft X-ray selected 
sources in the ROSAT Lockman Deep Field
(Schmidt et al. 1998),
reaching a flux depth of $f_{0.5-2} \approx 
10^{-15}$ $\rm erg~cm^{-2}~s^{-1}$,
65 and 15 per cent are
broad-line and narrow-line AGN respectively. 
A small contamination (6 per cent) by stars is also expected
(Schmidt et al. 1998), but since they are randomly distributed 
over the sky their effect 
would be to dilute somewhat the measured correlation signal.
From the work of Woods \& Fahlman (1997) we can deduce that
for a stellar contamination of 6 per cent 
a reduction in the observed correlation signal by $\sim$13 per cent 
should be expected (correcting for this reduces our $r_{\circ}$
values, derived in section 4 by only $\sim$5\%).
In the ROSAT Lockman Hole Survey there is also a small fraction of
galaxy groups. As these may be more strongly  clustered than galaxies,
and possibly AGN, they may increase marginally the overall signal. 
For this reason we have excluded the nine extended sources, which we
have found in our XMM observations, from the subsequent analysis.

The differential X-ray source counts in the 0.5-2 and 0.5-8\,keV
spectral bands are shown in Figure \ref{fig_lognlogs} and are
compared with the best fit relations of Baldi et al. (2002; soft band)
and Manners et al. (2003; total band). In the 0.5-2\,keV band there is
good agreement between our results and the Baldi et al. (2002) double 
power-law best fit to the number counts. The  Manners et al. (2003)
best fit is derived for sources in the flux range $f(\rm 0.5-8\,keV)
\rm = 10^{-15} - 8 \times 10^{-14} \, erg \, s^{-1} \,
cm^{-2}$. Although our $dN / dS$ is in good agreement with their  
results in the above flux range, at brighter fluxes the surface 
density of X-ray sources is lower than the extrapolated Manners et
al. (2003) relation. This suggests that a double power-law is required to
fit the 0.5-8\,keV $dN / dS$ over the flux range $\rm 10^{-15} -
10^{-12} \, erg \, s^{-1} \, cm^{-2}$. We therefore
adopt a double power-law  of the form: 
$$\log\frac{dN}{dS}=\left\{
\begin{array}{cc}
 A_1 + B_1 \times \log f_X & \: f_X<f_X^c \\
 A_2 + B_2 \times \log f_X & \: f_X\ge f_X^c\\
\end{array}
\right.$$
\noindent
where $A_2=A_1+(B_1-B_2)\times f_X^c$,  $f_X^c$ is the flux at the break.
%and $f_X$ is in units of $\rm erg\,s^{-1}\,cm^{-2}$. 
We estimate $f_X^c(\rm 0.5 - 8 \,keV)\approx \rm 6 \times 10^{-14} \, erg \,
s^{-1} \, cm^{-2}$,  $A_1=-8.9\pm2.2$, $B_1=-1.8\pm0.2$ and
$B_2=-2.3\pm0.1$. Our best-fit double power-law relation is shown
in Figure \ref{fig_lognlogs}. 

\section{Two-point correlation function analysis}
The two-point angular correlation function, $w(\theta)$, is defined as
the joint  probability of finding sources separated by an angle
$\theta$. For a random distribution of sources  $w(\theta)=0$ and
therefore, the  angular correlation function provides a measure of
galaxy density excess over that expected for  a random distribution. 
In this paper we use  the estimator  described by Efstathiou et
al. (1991) 

\begin{equation}
w(\theta)=f\frac{N_{DD}}{N_{DR}}-1,
\end{equation}

\noindent 
with the uncertainty in  $w(\theta)$ is estimated from the relation

\begin{equation}                        
\sigma_{w}=\sqrt{(1+w(\theta))/N_{DR}},
\end{equation}

\noindent 
where $N_{DD}$ is the number of data-data pairs in the interval
$[\theta-\Delta \theta,\theta+\Delta \theta]$ and $N_{DR}$ is the
number of data-random pairs  for a given separation. In the above
relation $f$ is the normalization factor $f = 2 N_R /(N_D-1)$ with
$N_D$ and $N_R$ being the total number of data and random points
respectively. 
For each XMM pointing we produce 100 Monte Carlo random 
catalogues having the same number of points as the real data 
 which also account for the sensitivity variations across 
the surveyed area (see
section 2). % which might affect the $w(\theta)$ estimate. 
Furthermore since the flux threshold for source detection depends on
the off-axis angle from the center of each of the XMM-{\it Newton} 
pointing, the sensitivity maps are used to discard random points in
less sensitive areas. This is
accomplished by assigning a flux to each random point using the
differential source counts plotted in Figure \ref{fig_lognlogs}. If
that flux is less than 5 times the local {\em rms}  noise at the
position of the random point (assuming Poisson statistics for the
background) this is excluded from the random data-set. We have
verified that our random simulations reproduce 
both the off-axis sensitivity of the detector as well as 
the individual field $\log N - \log S$.

Using the methods described above we estimate $w(\theta)$ in
logarithmic intervals with $\delta \log \theta\simeq 0.05$. 
For both samples we estimate $w(\theta<150^{''})\simeq 0.11\pm
0.03$ corresponding to a statistically significant signal at the 
$\approx 3.5\sigma$ confidence level (Poisson statistics). 
We now fit the measured  correlation function assuming a power-law of the
form $w(\theta)= (\theta_{\circ} /  \theta) ^ {\gamma-1}$, fixing
$\gamma$ to 1.8. We use a standard $\chi^{2}$ minimization procedure:
\be
\chi^{2}(\theta_{\circ})=\sum_{i=1}^{n} \left[ \frac{w_{\rm XMM}(\theta^{i})-
(\theta_{\circ}/\theta^{i})^{\gamma-1}}
{\sigma^{i}}\right]^{2} 
\ee 
with each point weighted by its error ($\sigma^{i}$). Note, that 
the fitting is performed for
angular separations in the range 40--1000\,arcsec. We also note that
our results  are insensitive to both the upper cutoff limit in
$\theta$ and the angular binning (for more than 10 bins)
used to estimate  $w(\theta)$. Therefore,  
the best fit parameters for both the soft and the total band
sub-samples are: 
$\theta_{\circ}=9.3\pm 1.9$ and $\theta_{\circ}=9.0\pm 1.7$ arcsec's
respectively. Note that the errors
correspond to $1\sigma$ ($\Delta \chi^{2}=1.00$) uncertainties, 
which are estimated using the  variation of $\Delta \chi^{2}=
\chi^{2}(\theta_{\circ})-\chi_{\rm min}^{2}(\theta_{\circ})$ 
[$\chi_{\rm min}^{2}$ is the absolute  minimum value of the $\chi^{2}$].
However, these raw values should be corrected for two 
possible bias presented below.

\tabcolsep 9pt
\begin{table*}
\caption{Angular correlation function analysis results. 
The columns are as: X-ray sub-sample, number of objects in the
sub-sample, the corresponding 
angular correlation length, the reduced $\chi^{2}$, the $\chi^{2}$ 
probabilities
and the correlation signal between 0-150 arcsec. 
The errors represent $1\sigma$ uncertainties.} 

\tabcolsep 6pt
\begin{tabular}{cccccc} 
\hline
X-ray band& No. of sources& $\theta_{\circ}$(arcsec)& 
$\chi^{2}/{\rm dof}$& $P_{\chi^{2}}$& $w(\theta <150^{''})$ \\ \hline 
 0.5-8\,keV  &  462 &  $10.8 \pm 1.7$ & 1.50& 0.10& $0.114\pm 0.037$\\
 0.5-2\,keV  &  432 &  $10.4 \pm 1.9$& 1.10& 0.35& $0.105\pm 0.035$\\
% (2-8)keV  &  177 &  $22.2^{+9.4}_{-8.6}$&0.56\\ \\
% 0.5-8\,keV-NGP  &  268 &  $8.2 \pm 3.0$& 1.40& 0.14& $0.09\pm 0.05$\\
% 0.5-8\,keV-SGP  &  207 &  $20.0\pm 5.8$& 1.90& 0.04& $0.18\pm 0.05$\\ \\
% 0.5-2\,keV-NGP  &  264 &  $6.0\pm 4.0$& 0.80& 0.70&$0.05\pm 0.05$\\
% 0.5-2\,keV-SGP  &  188 &  $29.0\pm 6.7$& 1.21& 0.27& $0.22\pm 0.05$\\
\hline
\end{tabular}
\end{table*}

\subsection{Integral constraint}
When calculating the angular correlation function from a bounded
region of solid angle $\Omega$, corresponding to the area of the
observed field, the background projected local density of sources is
$N_{s}/\Omega$ (where $N_{s}$ is the number of objects
brighter than a given flux limit). However, this is an
overestimation of the true underlying mean surface density, because of
the positive correlation between galaxies at small separations,
balanced by negative values of $w(\theta)$ at larger separations. This
bias, known  as the integral constraint, has the effect of reducing
the amplitude of the correlation function by  
 
\begin{equation}\label{eq_ic}
\omega_{\Omega}=\frac{1}{\Omega ^{2}} \int{\int{w(\theta) d\Omega_{1}
d\Omega_{2}}}.   
\end{equation} 
 
\noindent 
Clearly, evaluating $\omega_{\Omega}$ necessitates {\em a priori}
knowledge of the angular correlation function. A tentative value of
$\omega_{\Omega}$ using a range of $w(\theta)$ by varying within
1$\sigma$ our results is: $\omega_{\Omega}\simeq 0.01$. 

Adding $\omega_{\Omega}$ to each bin of
our raw $w(\theta)$ the integral constraint has a small 
but not negligible effect on
the estimated correlation lengths. Indeed, for the 0.5-2\,keV  band
repeating the fittings using $\omega_{\Omega} \simeq 0.01$
we find $\theta_{\circ} \simeq 10.7 \pm 1.9$\,arcsec
and $\theta_{\circ} \simeq 11.1 \pm 1.7$\,arcsec
for the soft and the total band respectively. 

\subsection{Amplification bias}
Another bias that may affect the measured angular correlation
function of our X-ray sources  is the {\em amplification bias}
(e.g. Vikhlinin \& Forman 1995). The original quantification of this
effect can be traced back to Kaiser's (1984) work which showed that
smoothing of the galaxy distribution using a Gaussian kernel with
size similar or larger to the correlation length of the underlying
galaxy distribution increases the correlation function of the
resulting density peaks compared to that of the underlying
galaxies. Furthermore, the larger the smoothing radius the higher the 
amplitude of the correlation function of the resulting density peaks.  

In the present analysis we are faced with a similar situation  
since the Point Spread Function (PSF) Full Width Half Maximum (FWHM)
of the XMM-{\it Newton} detector is of the same order of magnitude
($\sim 6$ arcsec) with the measured angular correlation length
($\theta_o\sim 11$ arcsec). X-ray Sources separated by less
than $\sim  6$ arcsec will be observed as a single
object. This is in effect a  smoothing process, similar to that of the
Kaiser's study, with smoothing radius roughly equal to the XMM-{\it
Newton} PSF size. Vikhlinin \& Forman (1995) studied the clustering
properties of X-ray sources detected on ROSAT archival data and found
that their measured $w(\theta)$ was severely affected by the
amplification bias due to the large FWHM of the ROSAT PSF. In our case
we expect  significantly less problems since the XMM-{\it Newton} PSF
size is smaller than that of the ROSAT detector.

We quantify this effect  using an approach that is similar to that of
Vikhlinin \& Forman (1995). These authors used the Soneira \& Peebles  
(1978) algorithm to construct correlated point processes with a
variety of in-build correlation amplitudes. Then using  a smoothing
window with the size of the ROSAT PSF they were able to determine
that their measured  $w(\theta)$ was artificially enhanced by a factor
of $\sim 2.85$.  

Our method is based on the concept that the galaxy correlation
function due to its power-law nature could be considered a
fractal. Therefore one can shift the amplitude of $w(\theta)$ at
different scales keeping
its slope fixed.  This simplifies our study since we do not need to
construct different correlated point process having a particular
correlation length. Any correlated point processes that is described
by power-law with the required exponent can be scaled to have a specific
correlation length one wishes. For our study we use the publically
available $\Lambda$CDM Hubble volume cluster distribution which has a
well defined power-law correlation function with an exponent $\gamma
\simeq 1.8$ (Frenk et al 2000).   

Lets assume that the angular correlation function of the
model catalogue above has a correlation length $\theta_{o,{\rm
c}}$ while, the true (unaffected from the amplification bias)
correlation length of the X-ray point sources is $\theta_{o,x}$. 
We can translate the angular scale of the model 
correlation function to that of the XMM correlations by multiplying
the former scale by the factor:
$$ f=\theta_{o,x}/\theta_{o,{\rm  c}}\;.$$ 
We can now simulate the effect of XMM PSF smoothing on the scaled model
correlation function by merging all the model pairs
with separations less than the  PSF FWHM (i.e. $\sim 
6$ arcsec for the XMM-{\it Newton}) and then fit the model
angular correlation function to obtain the best fit 
angular correlation length-scale and
compare it to that of the XMM point source data.

However, since we do not know the value of $\theta_{o,x}$ but it is rather
the value that we seek to find from our analysis, we apply an
iterative procedure by which we change the value
of $\theta_{o,x}$, and thus of $f$, until the
resulting scaled model correlation function (i.e. after smoothing) 
has an angular correlation length equal to that of the raw XMM point 
source correlations.

The previous analysis shows that our XMM-{\it Newton} observations are
only marginally affected by the amplification bias. The true
underlying angular correlation length of the X-ray population is
overestimated by $\sim$3-4 per cent. Therefore, we conclude 
that the corrected (free of the amplification bias) correlation length 
of the XMM-{\it Newton} soft X-ray sources is about $\theta_o\simeq
10.4$ arcsec.

We validate the above procedure by recalculating
the amplification bias for the case of ROSAT observations. We adopt 
a correlation length (free from amplification bias) of $\theta_{o,x} =
4$ arcsec (Vikhlinin \& Forman 1995) and an effective
smoothing scale of $\sim 20$ arcsec similar to  the ROSAT PSF
FWHM. Our procedure gives that the expected amplified correlation
length of the ROSAT sources is $\sim 10.5$ arcsec a factor of
$\sim 2.7$ higher than the true value, in excellent agreement with the   
Vikhlinin \& Forman (1995) analysis. We are therefore confident that
our method and results are robust. 

\subsection{The final angular correlation length}
After taking into account the corrections described above,
we present the corrected angular correlations in Figure 2
for the soft and total spectral bands. 
%In the insert of Figure 2 we present the angular correlation function
%using a linear scaling.
The best fit parameters for both sub-samples are presented in Table 1.
%$$\theta_{\circ}^{\rm soft} \simeq 10.4 \pm 2.9 \;\;\; {\rm arcmin}$$ 
%and
%$$\theta_{\circ}^{\rm total} \simeq 10.8 \pm 2.7 \;\;\; {\rm arcmin}$$ 
%As a further test we have repeated the fitting procedure but now
%leaving $\gamma$ as a free parameter. The corresponding results are:
%$$\theta_{\circ}^{\rm Soft} \simeq 29.4 \pm 12 \;\;\;{\rm with}\;\;\;
%\gamma=2.1\pm 0.3$$ 
%and
%$$\theta_{\circ}^{\rm Total} \simeq 25.1 \pm 10\;\;\;{\rm with}\;\;\;
%\gamma=2.1\pm 0.2\;\;.$$ 
\begin{figure}
\mbox{\epsfxsize=8.3cm \epsffile{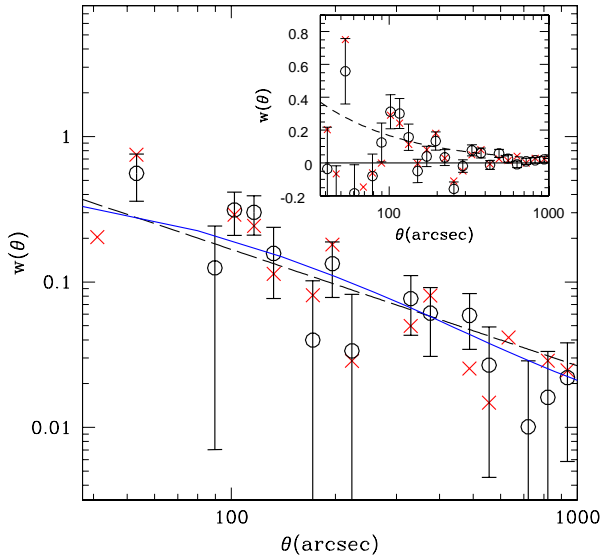}}
\caption{The two-point angular correlation function 
for the soft (open points) and total (crosses) bands respectively. The 
dashed line represent the best-fit power law 
$w(\theta)=(\theta_{\circ}/\theta)^{0.8}$ for the soft (0.5-2)keV
band (see parameters in Table 1), while the continuous line represents
the best fit $\Lambda$CDM ($\Omega_{\Lambda}=0.7$ and $h=0.7$) 
model in the framework of Basilakos \& Plionis (2001) 
biasing model.
%We plot only the best-fit and the errorbars for the soft band in 
%order to avoid confusion.
In the insert we present the measured 
correlation function in linear scaling.} 
\end{figure}
%ig
Our results are higher than those of Vikhlinin et al. 
 (1995) who derive the angular correlation function 
 from ROSAT pointed observations which have a comparable 
 effective flux limit with our XMM pointings. 
 The above authors find an angular clustering length 
 of $\theta_o\sim 10$ arcsec, which reduces to 4 arcsec 
 after correction for the amplification bias.    
 Comparison with the results of Basilakos et al. (2004)
shows that the hard band sources are more strongly  
clustered, at least  on angular projection, ($\theta_{\circ}^{\rm hard}
\simeq 22 \pm 9$ for $\gamma=1.8$) compared to the soft band sources.
%ig

\section{The spatial correlation length of the XMM soft sources}
\subsection{Inverting Limber's equation}
%The $r_{\circ}-\theta_{\circ}$ relation}
The spatial correlation function can be modeled as (de Zotti et al. 1990)
\begin{equation}\label{eq:dezotti}
\xi(r,z)=(r/r_{\circ})^{-\gamma}\times (1+z)^{-(3+\epsilon)}\;\;,
\end{equation} 
where $\epsilon$ parametrizes the type of clustering evolution.
If $\epsilon=\gamma-3$ (ie., $\epsilon=-1.2$ for $\gamma=1.8$), 
the clustering is constant in comoving coordinates (comoving clustering),
which means that the amplitude of the correlation function
remains fixed with redshift in comoving coordinates as the galaxy pair 
expands together with the Universal expansion.
Alternatively, in the $\epsilon=-3$ model 
the clustering is constant in physical coordinates, while 
$\epsilon=0$ reflects the {\em stable} clustering model (eg. de Zotti et
al. 1990).

We can relate the amplitude $\theta_{\circ}$ in two 
dimensions to the corresponding three dimensional one, $r_{\circ}$,
using Limber's integral equation (cf. Peebles 1993). For
example, in the case of a spatially flat Universe,
Limber equation can be written as
\be 
w(\theta)=2\frac{\int_{0}^{\infty} \int_{0}^{\infty} x^{4} 
\phi^{2}(x) \xi(r,z) {\rm d}x {\rm d}u}
{[\int_{0}^{\infty} x^{2} \phi(x){\rm d}x]^{2}} \;\; , 
\ee
where $\phi(x)$ is the selection function (the probability 
that a source at a distance $x$ is detected in the survey) and 
$x$ is the proper distance related to the redshift through 
\be
x(z)=\frac{c}{H_{\circ}} \int_{0}^{z} \frac{{\rm d}t}{E(t)}\;\; ,
\ee
with 
\be 
E(z)=[\Omega_{\rm m}(1+z)^{3}+\Omega_{\Lambda}]^{1/2} 
\ee
(see Peebles 1993). The number of objects in the given 
survey with a solid angle $\Omega_{s}$ and within 
the shell $(z,z+{\rm d}z)$ is:
\be
\frac{{\rm d}N}{{\rm d}z}=\Omega_{s}
x^{2}\phi(x)\left(\frac{c}{H_{\circ}}\right) E^{-1}(z)\;\;.
\ee
Therefore, combining the above system of 
equations, the expression for $w(\theta)$ satisfies the form 

\begin{equation}\label{eq:angu}
w(\theta)=2\frac{H_{\circ}}{c} \int_{0}^{\infty} 
\left(\frac{1}{N}\frac{{\rm d}N}{{\rm d}z} \right)^{2}E(z){\rm d}z 
\int_{0}^{\infty} \xi(r,z) {\rm d}u
\end{equation} 
Note that, the physical separation between two sources, 
separated by an angle $\theta$ considering 
the small angle approximation, is given by:
\be
r \simeq \frac{1}{(1+z)} \left( u^{2}+x^{2}\theta^{2} \right)^{1/2} \;\; .
\ee
Using eq.(\ref{eq:dezotti}) and eq.(\ref{eq:angu}) we obtain:
\be\label{eq:invert}
\theta_{\circ}^{\gamma-1}=H_{\gamma}
\left(\frac{r_{\circ}^{\gamma} H_\circ}{c}\right) 
\int_{0}^{\infty} \left( \frac{1}{N}\frac{{\rm d}N}{{\rm d}z}
\right)^{2} \frac{E(z) (1+z)^{-3-\epsilon+\gamma}}{x^{\gamma-1}(z)}  
{\rm d}z \;,
\ee
where $H_{\gamma}=\Gamma(\frac{1}{2})
\Gamma(\frac{\gamma-1}{2})/\Gamma
(\frac{\gamma}{2})$.  

In order to perform the inversion we still need to determine 
the source redshift distribution ${\rm d}N/{\rm d}z$. Since we have no
unbiased redshift information for our sources we can resort to a
measure of ${\rm d}N/{\rm d}z$ using an estimate of their luminosity function.
In flux-limited samples, there is a degradation of
sampling as a function of distance from the observer (codified by the
so called {\em selection function}). The latter also depends on the evolution
of the source luminosity function but it is independent of the
cosmological model, used in the derivation of the luminosity function.
Thus for our X-ray sources the selection function can be written as: 
\be 
\phi(x)=\int_{L_{\rm min}(z)}^{\infty} \Phi(L_{x},z) {\rm d}L \;\;,
\ee
where $\Phi(L_{x},z)$ is their redshift dependent luminosity function.
In this work we used the soft band luminosity
functions of Miyaji, Hasinger \& Schmidt (2000) and of Boyle et al. (1993).
We also use different models for the evolution of the soft
X-ray sources: a pure luminosity evolution (PLE) or the more realistic
luminosity dependent density evolution (LDDE; Miyaji et al 2000). In
Fig. 3 we present the expected redshift distributions of the
soft X-ray sources for three different luminosity functions and
evolution models. The LDDE model predicts a redshift
distribution shifted to much larger redshifts with a 
median redshift of $\bar{z} \simeq 1.2$ (see also Table 2) comparing with
both the Boyle et al (1993) and
Miyaji et al (2000) luminosity functions with pure luminosity
evolution.
It is very interesting that the source 
redshift distribution of the ROSAT Lochman Deep field (Schmidt et
al. 1998), albeit having a flux limit slightly lower than of our survey, traces
quite well the LDDE predictions (see histogram in
Fig. 3), a fact that supports this luminosity function 
evolutionary model. To quantify this claim we have performed a $\chi^{2}$
test between the observed and theoretical redshift distributions and
found that the probability of consistency is 0.45 ($\chi^{2}/{\rm 
df}=0.97$), $<10^{-6}$ ($\chi^{2}/{\rm df}=8.5$) and 0.04
($\chi^{2}/{\rm df}=2.1$) for the LDDE, PLE (Miyaji) and the PLE
(Boyle) models, respectively.

\begin{figure}\label{fig:reds}
\mbox{\epsfxsize=9cm \epsffile{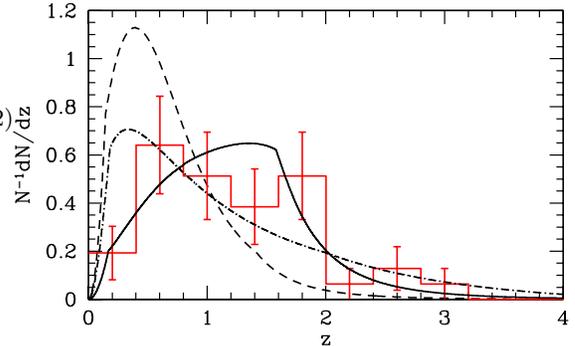}}
\caption{The redshift selection function
for three different luminosity function models:
(a) Miyaji et al (2000) with PLE (dashed line), (b)
Miyaji et al (2000) with LDDE (continuous line)
and (c) Boyle et al. (1993) with PLE (dot-dashed line). The histogram
corresponds to the distribution of the Schmidt et al (1998) X-ray 
sources of the ROSAT Lochman Deep Field.}
\end{figure}

\begin{table*}
\caption[]{The soft X-ray sources 
correlation length ($r_{\circ}$ in $h^{-1}$ Mpc)
for different clustering models ($\epsilon$) and 
for the different luminosity functions and evolution models. The last
column indicates the predicted median 
redshift, from the specific luminosity function
used. The bold letters delineate the preferred cosmological model and
the most updated luminosity function.}
\vspace{0.2cm}
\tabcolsep 5pt
\begin{tabular}{cccccc} \hline
LF & Evol. Model & ($\Omega_{m},\Omega_{\Lambda}$) &
$r_{\circ}$ $(\epsilon=-1.2)$&
$r_{\circ}$ $(\epsilon=-3)$ & $\bar{z}$ \\ \hline
Boyle & No evol. & (1,0)   & 7.9$\pm 0.6$    & 5.4$\pm 0.4$ &0.50 \\
Miyaji  & No evol. & (1,0)   & 6.5$\pm 0.5$ & 4.9$\pm 0.4$&0.37 \\
Boyle & PLE      & (1,0)   & 12.0$\pm 1.0$   & 6.3 $\pm 0.5$&0.92 \\
Miyaji  & PLE & (1,0) & 8.8$\pm 0.7$ &$5.7\pm 0.4$ & 0.58 \\
{\bf Miyaji} & {\bf LDDE} & {\bf (0.3,0.7)} & ${\bf 16.4 \pm 1.3}$ & ${\bf
  7.5 \pm 0.6}$&{\bf 1.19} \\
Miyaji  & LDDE     & (1,0)   & $11.2 \pm 0.9$ & 5.0 $\pm 0.4$&1.19 \\ \hline
\end{tabular}
\end{table*}

\subsection{Results}
Using eq.(\ref{eq:invert}), the LDDE luminosity evolution model and $\epsilon=-1.2$ 
we find within the concordance cosmological model a soft band 
correlation length of $r_{\circ}\simeq 16.4\pm 1.3 \;h^{-1}$ Mpc.
This value is comparable to that of Extremely 
Red Objects (EROs), luminous radio  sources (Roche, Dunlop \& 
Almaini 2003; Overzier et al. 2003; R\"{o}ttgering et al. 2003) and 
hard X-ray sources (Basilakos et al. 2004)
which are found to be in the range $r_{\circ} \simeq 12 - 19 \;
h^{-1}$ Mpc. It is interesting to mention that also 
radio sources which contain an AGN
show strong clustering ($r_{\circ} \simeq 11\;
h^{-1}$ Mpc), while the opposite is true for the case
of radio sources showing no AGN activity (Magliocchetti et al. 2004). 

However, our previously derived $r_{\circ}$ value is significantly larger than those 
derived from optical AGN surveys: $r_{\circ} \simeq 5.4 - 8.6 \; h^{-1}$ Mpc
(Croom \& Shanks 1996; La Franca et al. 1998; Akylas et al. 2000; 
Croom et al. 2002; Grazian et al. 2004) 
as well as from the recent X-ray selected sample of
Mullis et al. (2004) who find $r_{\circ}\simeq 7.4$ $h^{-1}$ Mpc.
We can push our inverted $r_{\circ}$ values to approximate closely 
the latter results only if we use the constant in physical coordinates
clustering evolution model 
($\epsilon=-3$), in which case we obtain $r_{\circ} \simeq 7.5\pm 0.6 
\;h^{-1}$ Mpc, which is in excellent agreement with the 
Mullis et al. (2004) results.
Note that earlier {\sc ROSAT} soft X-ray clustering 
results of Carrera et al (1998) found a weaker clustering, 
with their upper limit of the linear clustering evolution model being 
 marginally consistent with our $\epsilon=-3$ results. 

In Table 2, we list the values of the correlation
length, $r_{\circ}$, resulting from Limber's inversion for different
luminosity function, evolution models as well as for
different cosmological models. We can attempt to disentangle the
different sources of the apparent $r_{\circ}$ differences.
Firstly, comparing the LDDE model between the Einstein de Sitter and
the concordance  ($\Omega_{\rm m}=1-\Omega_{\Lambda}=0.3$)
Cosmological models it becomes evident that the effect of
moving from the former to the latter model increases by
$\sim 50\%$ the value of $r_{\circ}$ (for both $\epsilon$ cases). 
Note also that within the EdS cosmological model moving from the LDDE to
the PLE luminosity evolution models decreases the value of 
$r_{\circ}$ by $\sim 20\%$ for $\epsilon=-1.2$, while for the
$\epsilon=-3$ case there is no significant difference between the two
luminosity evolution models. Therefore, although we do not
have the PLE luminosity model parameters for the concordance cosmological
model we may expect similar changes as before, which implies that within this
cosmological and luminosity evolution (PLE) models
we would obtain $r_{\circ}\sim 12.5$ and 7 $h^{-1}$ Mpc for the
$\epsilon=-1.2$ and $\epsilon=-3$ models, respectively.

Also note that the change of the
luminosity function model and thus of the redshift selection function,
is always accompanied by a change of the
median redshift of the corresponding redshift distribution.

We can attempt to parametrize the different luminosity model effects
on the determination of $r_{\circ}$ by investigating its dependence
on the median redshift of the source distribution 
as well as on the cosmological model. To do so we use  
a parametrized, by the characteristic redshift $\bar{z}$,
analytical selection function, given by Baugh (1996):
\be\label{eq:param}
\frac{{\rm d}N}{{\rm d}z} \propto z^{2} 
{\rm exp}\left[ -\left(\frac{z}{z_{c}} \right)^{3/2} \right] \;\;,
\ee
where $\bar{z}=\sqrt{2}z_{c}$ is the median redshift.
Although this formula has been derived from the distribution 
of optical galaxies while the redshift distribution 
of X-ray sources maybe different we find that at least for the case of
the Miyaji et al (2000) luminosity function provides absolutely
consistent results. For example, inserting eq.~(\ref{eq:param}) in
eq.~(\ref{eq:invert}) with $\bar{z}\simeq 1.19$ and $\epsilon=-1.2$  
we find for the LDDE model
$r_{\circ}\simeq 16.3\pm 2.0 \;h^{-1}$ Mpc and 
$r_{\circ}\simeq 11.0\pm 1.5 \;h^{-1}$ Mpc 
for the concordance 
and Einstein de Sitter models, respectively, which are in excellent
agreement with the direct LDDE results (see Table 2). Similar 
consistency is found also for the other models presented in 
Table 2 (except for the models based on Boyle's luminosity function
which is due to their significant contributions from very large redshifts).
%These results leads us to believe that indeed a valid parametrization
%of the redshift distribution resulting from a 
In Fig. 4, we show the dependence of the derived $r_{\circ}$ on
the median redshift of the source distribution for the two different
cosmologies and two clustering evolution models. 
We also plot our direct results (using the different Miyaji et al.
luminosity function models that provide different ${\bar z}$) of Table
2. The excellent consistency is evident which makes us confident of
our results.
\begin{figure}
\mbox{\epsfxsize=8.3cm \epsffile{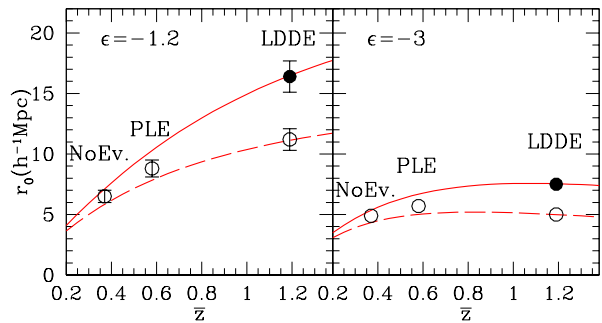}}
\caption{Comparison of the expected increase of $r_{\circ}$ as a function
of the median redshift of the source redshift distribution for a
concordance (continuous line) and an Einstein de Sitter model
respectively, using the parametric selection function model of Baugh (1996). The 
results of Table 2 based on the Miyaji et al (2000) luminosity
functions are plotted (filled circles for the concordance model 
and open for the Einstein de Sitter model).
}
\end{figure}
Guided by Fig. 4 we can deduce that the PLE model of the Miyaji et al
(2000) luminosity function would provide $r_{\circ} \simeq 12$ and 7
$h^{-1}$ Mpc within the concordance cosmological model for the
$\epsilon=-1.2$ and $\epsilon=-3$ clustering models respectively.

\section{The XMM sources cosmological bias}
Within the framework of
linear biasing (cf. Kaiser 1984; Benson et al. 2000), the mass-tracer 
and dark-matter spatial correlations, at some redshift $z$, are related by:
\be
\label{eq:spat}
\xi(r,z)=\xi_{\rm DM}(r,z)b^{2}(z) \;\;, 
\ee
where $b(z)$ is the bias evolution function. 

We can quantify the evolution of clustering with epoch
presenting the spatial correlation function of the mass 
$\xi_{\rm DM}(r,z)$ as the Fourier transform of the 
spatial power spectrum $P(k)$:
\be
\xi_{\rm DM}(r,z)=
(1+z)^{-(3+\epsilon)}
\frac{1}{2\pi^{2}}\int_{0}^{\infty} k^{2}P(k) 
\frac{{\rm sin}(kr)}{kr}{\rm d}k \;,
\ee
where $k$ is the comoving wavenumber. 
Furthermore, the predicted spatial correlation function of the
X-ray sources can be written as:
\be
\label{eq:spat1}
\xi(r,z)=\frac{R(z)}{2\pi^{2}}\int_{0}^{\infty} k^{2}P(k) 
\frac{{\rm sin}(kr)}{kr}{\rm d}k \;\;,
\ee
where
\be
\label{eq:spat2}
R(z)=(1+z)^{-(3+\epsilon)}b^{2}(z) \;\;\;.
\ee

As for the power spectrum of our CDM models, we use 
$P(k) \approx k^{n}T^{2}(k)$ with scale-invariant ($n=1$) primeval 
inflationary fluctuations and $T(k)$ the CDM transfer function.
In particular, we use the transfer function parameterization as in
Bardeen et al. (1986), with the corrections given approximately
by Sugiyama (1995). 
Note that we also use the
non-linear corrections introduced by Peacock \& Dodds (1994).  

\subsection{Bias Evolution}
The concept of biasing between different classes of extragalactic objects 
and the background matter distribution was put forward by Kaiser (1984)
and Bardeen et al. (1986) 
in order to explain the higher amplitude of the 2-point correlation function 
of clusters of galaxies with respect to that of galaxies themselves.

The deterministic and linear nature of 
biasing has been challenged (cf. Bagla 1998; Dekel \& Lahav 1999) and indeed 
on small scales ($r< 10h^{-1}$Mpc) there are 
significant deviations from $b(r)=const$. 
Despite this, the linear biasing assumption is still a useful first order
approximation which, due to its simplicity, it is used in most studies
of large scale clustering (cf. Magliocchetti et al. 1999). 
In this paper however, 
we will work within the paradigm of linear and 
scale-independent bias. Based on different assumptions a number 
of bias evolution models have been
proposed (eg.  Nusser \& Davis 1994; Fry 1996; Mo \& White 1996;
Matarrese et al. 1997; Tegmark \& Peebles 1998; Bagla 1998; 
Plionis \& Basilakos 2001). However, here we will discuss two that
have been shown to describe relatively well the evolution even beyond 
$z\sim1$.

\begin{itemize}

\item {\em Merging Bias Model} (hereafter B2): 
Mo \& White (1996) have developed a model for 
the evolution of the the so-called correlation bias, 
using the Press-Schechter formalism.
Utilizing a similar formalism, Matarrese et al. (1997) extended 
the Mo \& White (1996)
results to include the effects of different mass scales (see also 
Moscardini et al. 1998; Bagla 1998; Catelan et al. 1998; Magliocchetti
et al. 2000). In this case we have that
\be
b_{\rm B2}(z)=0.41+\frac{(b_{\circ} - 0.41)}{D^{\beta}(z)} \;\;\; ,
\ee
with $\beta \simeq 1.8$.
Note that $D(z)$ is the linear growth rate of clustering (cf. Peebles 1993)
\footnote{$D(z)=(1+z)^{-1}$ for an Einstein-de Sitter Universe.} 
scaled to unity at the present time.

\item {\em Bias from Linear Perturbation Theory} (hereafter B3): 
Basilakos \& Plionis (2001, 2003), using linear
perturbation theory and the 
Friedmann-Lemaitre solutions of the cosmological
field equations have derived analytically the functional form
for the evolution of the 
linear bias factor, $b$, between the background matter and 
a mass-tracer fluctuation field.
For the case of a spatially flat $\Lambda$ cosmological model
($\Omega_{\rm m}+\Omega_{\Lambda}=1$), the bias 
evolution can be written as:

\be\label{eq:84}
b_{\rm B3}(z)={\cal A} E(z)+{\cal C}E(z)K(z)+1
\ee
with
\be
K(z)=\int_{1+z}^{\infty} \frac{y^{3}}
{[\Omega_{\rm m} y^{3}+\Omega_{\Lambda}]^{3/2}} {\rm d}y
\ee
or
\be
K(z)=(1+z)^{-1/2} 
F\left[\frac{1}{6},\frac{3}{2},\frac{7}{6},-\frac{\Omega_{\Lambda}}
{\Omega_{\rm m} (1+z)^{3}} \right]
\ee
where $F$ is the hyper-geometric function. 
Note that this approach gives a family of bias
curves, due to the fact that it has two unknown parameters, 
(the integration constants ${\cal A},{\cal C}$). 
Basilakos \& Plionis (2001, 2003) 
compared the B3 bias evolution model with other models as well
as with the HDF (Hubble Deep Field) biasing 
results (Arnouts et al. 2002; Malioccietti
1999), and found 
a very good consistency. In this work, for simplicity, we 
fix the value of ${\cal C}$ being $\simeq 0.004$, as was determined
in Basilakos \& Plionis (2003) from the 2dF galaxy correlation function. 
%%%Begining Spyros and then Manolis
It is evident that the bias factor at the present time
can be obtained from eq.(\ref{eq:84}) for $z=0$
\be
b_{\rm B3}(0)={\cal A}+{\cal C}K(0)+1
\ee
where we have used $E(0)=1$. Note that $K(0)\simeq 9.567$ for 
$\Omega_{\Lambda}=1-\Omega_{\rm m}=0.7$.

\begin{figure}
\mbox{\epsfxsize=8.3cm \epsffile{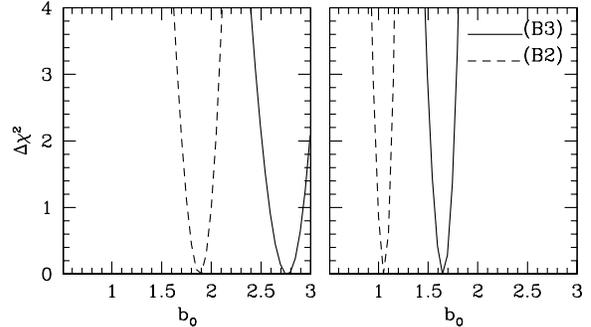}}
\caption{The variation of $\Delta \chi^{2}$ around the best
bias fit ($b_{\circ}$) using different clustering 
behaviours (left panel for $\epsilon=-1.2$ and right panel for 
$\epsilon=-3$). 
Note that the solid and dashed lines represent 
the bias from the linear perturbation  
(B3) and the merging (B2) bias models respectively.
}
\end{figure}

\begin{figure}
\mbox{\epsfxsize=8.3cm \epsffile{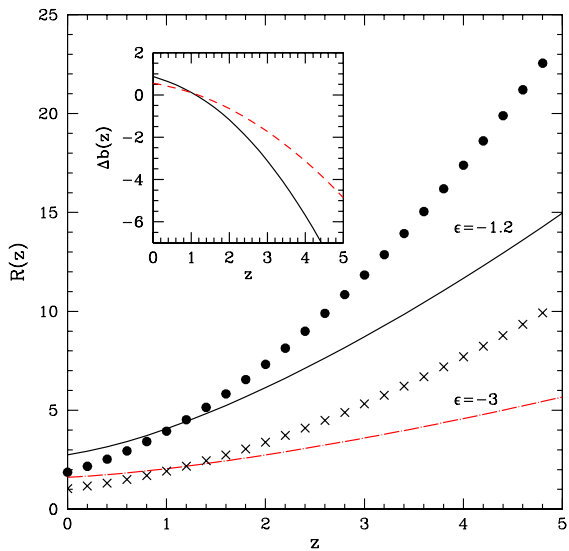}}
\caption{The function $R(z)=(1+z)^{-(3+\epsilon)}b^{2}(z)$ as a 
function of redshift.
The continuous line (B3 bias) and the filled point (B2 bias) types 
represent the $R(z)$ behaviour in the framework of a comoving clustering 
($\epsilon=-1.2$), while the dot dashed line (B3 bias) and the 
crosses (B2 bias) are based on a
clustering model which is constant in physical coordinates
($\epsilon=-3$).
In the insert we present the
difference, $\Delta b(z)=b_{\rm B3}(z)-b_{\rm B2}(z)$, 
between the (B3) and (B2) bias models as a function of $z$
for $\epsilon=-1.2$ (continuous line) and 
$\epsilon=-3$ (dashed line) respectively.
}
\end{figure}

\end{itemize}

\subsection{The bias at the present time $b_{\circ}$.}
Based on the luminosity dependent density evolution 
(LDDE; Miyaji et al 2000) we
quantify the bias factor at the present time $b_{\circ}$,
performing a standard $\chi^{2}$ 
minimization procedure between the measured 
correlation function for the soft band (0.5-2)keV
with that expected in our $\Lambda$CDM cosmological model,
\be
\chi^{2}(b_{\circ})=\sum_{i=1}^{n} \left[ \frac{w_{\rm XMM}(\theta^{i})-
w_{\rm model}(\theta^{i},b_{\circ})}
{\sigma^{i}}\right]^{2} \;\;.
\ee 
where $\sigma^{i}$ is the observed $w(\theta)$ uncertainty.

In Fig. 5 we present for the two bias and clustering evolution models 
the variation of 
$\Delta \chi^{2}=\chi^{2}(b_{\circ})-\chi^{2}_{\rm min}(b_{\circ})$ 
around the best $b_{\circ}$ fit, 
while in Table 3 we list the results of the 
corresponding fits for all the considered models.
The resulting present time bias is between: $b_{\circ}\simeq 1.05-1.90$ and 
$b_{\circ}\simeq 1.64-2.74$ for the B2 and B3 bias models, respectively.
Note that the theoretical ($\Lambda$CDM $+$ B3 bias model) 
fit to the measured soft X-ray source 
angular correlation function is presented as the solid line in Fig.2.
\begin{table}
\caption[]{Results of the predicted soft X-ray sources bias. 
The correspondence of the columns is as follows: bias and clustering
evolution models,
$b_{\circ}$ is the bias at the present time, the reduced
$\chi^{2}$ and the $\chi^{2}$ probabilities.
Errors of the fitted parameters represent $1\sigma$ uncertainties.}

\tabcolsep 9pt
\begin{tabular}{ccccc} 
\hline
Bias Model& $\epsilon$& $b_{\circ}$&
$\chi^{2}/{\rm dof}$& $P_{\chi^{2}}$
\\ \hline 
 B2  &   -1.2 &  $1.90\pm 0.10$&0.97&0.49\\
 B3  &   -1.2 &  $2.74\pm 0.15$&0.86&0.60\\ \\
 B2  &   -3.0 &  $1.05\pm 0.05$&0.97&0.49\\
 B3  &   -3.0 &  $1.64\pm 0.05$&0.84&0.63\\ \hline 
\end{tabular}
\end{table}

In order to understand better the effects of AGN clustering,
we present in Fig. 6 the quantity 
$R(z)$
%=(1+z)^{-(3+\epsilon)}b^{2}(z)$ 
(see eq.\ref{eq:spat2}) as a 
function of redshift for the concordance cosmological model and for
different bias evolution models. It is quite
obvious that the behaviour of the function $R(z)$ characterizes the  
clustering evolution with epoch; in general AGN clustering is a monotonically 
increasing function of redshift for both B2 and B3 biasing models.
Figure 6, for example, 
clearly shows that the bias at high redshifts has different 
values in the different clustering models. In particular, for the 
comoving clustering, $\epsilon=-1.2$ (continuous line for B3 and 
filled points for B2), 
the distribution of soft X-ray sources is 
strongly biased ($1.90 \mincir b_{\circ} \mincir 2.74$), as 
opposed to the less biased distribution 
($1.05 \mincir b_{\circ} \mincir 1.64$) in the $\epsilon=-3$ 
(dashed line for B3 and crosses for B2)
clustering model.       
This is to be expected, simply because the value 
$\epsilon=-3$ removes the $(1+z)$ dependence
from the $R(z)$ functional form and thus, produces
a lower corresponding correlation length 
(see Table 2),
in contrast with the comoving ($\epsilon=-1.2$) clustering case.
In other words, the higher or lower correlation length 
corresponds to a higher or lower bias at 
the present time respectively, being 
consistent with the hierarchical clustering 
scenario (cf. Magliocchetti et al. 1999).
Note, that the above predictions 
are in good agreement with those derived by 
Treyer et al. (1998), Carrera et al. (1998), Barcons et al. (2000) 
and Boughn \& Crittenden (2004), who have found $b_{\circ} \sim 1-2$.

Regarding the predictions of the two bias models, we present 
in the insert of Fig. 6 the 
difference, $\Delta b(z)=b_{\rm B3}(z)-b_{\rm B2}(z)$, 
between the B3 bias model and the Mataresse et al. (1997) model B2
as a function of redshift. The B2 bias evolves significantly more 
than the B3 model at relatively low redshifts ($z \le 1.0$), which
could be attributed to our assumption that the galaxy number density
is conserved in time. It is evident that merging processes,
not taken into account in the B3 model, 
are probably important in the evolution of clustering.

It is evident that the behaviour of the inverted X-ray source spatial
correlation function is sensitive to the different values of $\epsilon$ but 
there is also a strong dependence on the bias models that 
we have considered in our analysis.

We can attempt to select the most viable bias 
and the clustering evolution models by:
\begin{enumerate}
\item invoking the results of the local X-ray AGN 
clustering  of Akylas et al (2000) and Mullis et al. (2004), 
who find $r_{\circ}\simeq 6.5$ and 7.4 $h^{-1}$ Mpc, respectively
and 
\item noting that the local galaxy distribution, with a correlation
length $r_{\circ}\simeq 5 \; h^{-1}$ Mpc, is unbiased with respect
to the corresponding underline matter distribution (eg. Lahav et
al. 2002; Verde et al. 2002).
\end{enumerate} 
These two facts leads us to a local bias between the X-ray selected
AGN population and the underline matter distribution of:
$$b(0)=(r_{\circ, m}/r_{\circ, {\rm AGN}})^{-0.9} \sim (5/7)^{-0.9}\simeq
1.35 \;,$$ 
which is consistent only with the $\epsilon=-3$ model of
clustering evolution while it is in between the predictions of the
two bias models used.

\section{Conclusions}
%ig 
We have studied the angular clustering properties of the soft 
(0.5-e keV) X-ray point sources found in the {\it XMM-Newton}/2dF
survey. 
We find that there is a  strong ($3\sigma$) clustering
signal. Indeed, if the two point angular correlation function
is modeled as a power law, 
$w(\theta)=(\theta_{\circ}/\theta)^{0.8}$, then 
after correcting for the
integral constraint and the amplification bias 
the best-fitting angular clustering length is  
$\theta_{\circ}\simeq 10.4\pm 1.9$ arcsec.

Inverting Limber's equation and using the preferred
 luminosity dependent density evolution model for the luminosity
 function gives 
 $r_{\circ}\simeq 16$ and 7.5 $h^{-1}$ Mpc,
for the constant in comoving ($\epsilon=-1.2$) and in physical
 ($\epsilon=-3$) coordinates clustering evolution models, respectively. 
 In the former case, the values for the clustering length  
 are comparable with those of Extremely Red Objects (EROs) 
 and luminous radio sources, and are significantly higher 
 than those found from previous {\it ROSAT} surveys
 (e.g. Vikhlinin et al. 1995, Akylas et al. 2000, Carrera et al.
 1998; Mullis et al. 2004) and optical QSO surveys such as the 2QZ
 (Croom et al. 2002) and that of Grazian et al. (2004). 
However, we obtain a quite
 good agreement with the above surveys, only in the latter case
 of a clustering evolution model where the clustering length 
 remains constant in physical coordinates ($\epsilon=-3$).     

Comparing the measured angular correlation function  
for the soft band (0.5-2)keV X-ray sources
with the theoretical predictions of the preferred $\Lambda$CDM cosmological model 
($\Omega_{\rm m}=1-\Omega_{\Lambda}=0.3$) and two 
bias evolution models, we find that the present bias values is in the 
range of $1.9 \mincir b_{\circ} \mincir 2.7$ for the $\epsilon=-1.2$ model 
and $1.0 \mincir b_{\circ} \mincir 1.6$ for the $\epsilon=-3$ model.

\section* {Acknowledgments}
SB acknowledges the hospitality of the 
Astrophysics Department of INAOE where this work was completed. 
The $\Lambda$CDM simulation 
used in this paper was carried out by the Virgo 
Supercomputing Consortium using computers based at the Computing Centre of the
Max-Planck Society in Garching and at the Edinburgh parallel 
Computing Centre. The data are publicly
available at http://www.mpa-garching.mpg.de/NumCos.
We thank the referee, F.Carrera, for useful suggestions.

This work is jointly funded by the European Union
and the Greek Government in the framework of the program 'Promotion
of Excellence in Technological Development and Research', project
{\em 'X-ray Astrophysics with ESA's mission XMM'}. Furthermore, MP
acknowledges support by the Mexican Government grant No
CONACyT-2002-C01-39679.

\end{document}